\documentclass[aps,pra,preprint,groupedaddress,amsmath,amssymb,showpacs]{revtex4-1}
\usepackage{graphicx,amsmath,amssymb,epsfig}
\usepackage{dcolumn}
\usepackage{bm}
\usepackage{color}

\allowdisplaybreaks
\begin{document}

\renewcommand{\baselinestretch}{0.9}
\begin{center}
\LARGE \bf Low-loss Airy Surface Plasmon Polaritons
\\ %  Title
\parskip=0.5cm
%\\\vskip -0.01cm %  Title
%\parskip=0.5cm
%\normalsize (Invited Paper)\\
%\vskip 0.3cm
\renewcommand{\baselinestretch}{1.2}
\normalsize Qi Zhang$^{1}$, Chaohua Tan$^{1}$, Chao Hang$^{1}$, and Guoxiang Huang$^{1,*}$\\
\parskip=0.65cm
\small \sl{ State Key Laboratory of Precision Spectroscopy and Department of Physics,
East China Normal University, Shanghai 200062, China}\\
$^*$Corresponding author: gxhuang@phy.ecnu.edu.cn\\
%\rm Received March ?, 2015\vskip 0.05cm
\end{center}

\parskip=-0.6cm
\renewcommand{\baselinestretch}{0.92}

\begin{minipage}[t]{15cm}\normalsize \noindent \small
We propose a scheme to obtain a low-loss propagation of Airy surface plasmon polaritons (SPPs) along the interface between a dielectric and a negative-index metamaterial (NIMM). We show that, by using the transverse-magnetic mode and the related destructive interference effect between electric and magnetic absorption responses, the propagation loss of the Airy SPPs can be largely suppressed when the optical frequency is close to the lossless point of the NIMM. As a result, the Airy SPPs obtained in our scheme can propagate more than 6-time long distance than that in conventional dielectric-metal interfaces.
\\
\mbox{} \hspace{0.3cm}\sl OCIS codes: \rm 240.0240, 240.6680, 160.3918\\
\mbox{} \hspace{0.3cm}\sl doi: 10.3788/COL201412.000000.
\end{minipage}

\normalsize\parindent=0.3cm \noindent
\newpage

Dielectric-metal interfaces are known to support surface plasmon polaritons (SPPs), i.e. surface electromagnetic waves propagating along the planar interface between a metal and a dielectric material. These particular electromagnetic modes are sustained by collective electronic oscillations in the metal near the interface, and are essentially two-dimensional waves with field components decaying exponentially with distance from the interface. The fact that SPPs can overcome diffraction limit and localize light within a subwavelength volume makes them ideal tools for enhancing light-matter interaction [1].

\vskip 0.75cm

In recent years, much attention has been paid to the study of Airy light beams [2]. Unlike most types of light waves, Airy beams have the ability to resist diffraction, and can freely bend without requiring waveguiding structures or external potentials [3]. Up to now, Airy beams have been found in various fields of physics including optics [2,4-6], spin waves [7], plasma [8], and electron beams [9]. They also have a wide range of applications including trapping, guiding, manipulation of objects [10,11] and slow lights in cold atomic gases [12-13], signal processing [14], molecule diagnostics, and biosensing applications.

\vskip 0.75cm

Recently, Airy beams have been introduced theoretically by Salandrino and Christodoulides for controlling the SPPs along dielectric-metal interfaces [15]. This work opens an avenue for realizing nondiffracting SPPs, and stimulated a flourished experimental activities on Airy SPPs along dielectric-metal interfaces [16]. However, the Airy SPPs found up to now have a very short propagation distance due to the large Ohmic loss inherent in metals, which severely limits practical applications of the Airy SPPs.

\vskip 0.75cm

In the present work, we suggest a scheme to generate low-loss Airy SPPs, which propagate along the interface between a dielectric and a negative-index metamaterial (NIMM) [17-18]. By using the transverse-magnetic mode and the related destructive interference effect between electric and magnetic absorption responses,
we show that the propagation loss of the Airy SPPs can be suppressed significantly. As a result, the Airy SPPs obtained in our scheme can propagate more than 6-time long distance than that in conventional dielectric-metal interfaces, after taking the inherent diffraction effect into account.

\vskip 0.75cm

The system we consider consists of two superposed planar materials, i.e.
a dielectric and a NIMM (which could be replaced by metal for comparison), with a planar dielectric-NIMM interface, as shown in Fig.~1 (a). The NIMM in the lower half-plane ($x<0$) has permittivity $\varepsilon_1$
and permeability $\mu_1$ (which are frequency-dependent), and the dielectric in the upper half-plane ($x>0$) has permittivity $\varepsilon_2$ and permeability $\mu_2$
(which are taken as frequency-independent constants).
\begin{center}
\includegraphics[width=\columnwidth]{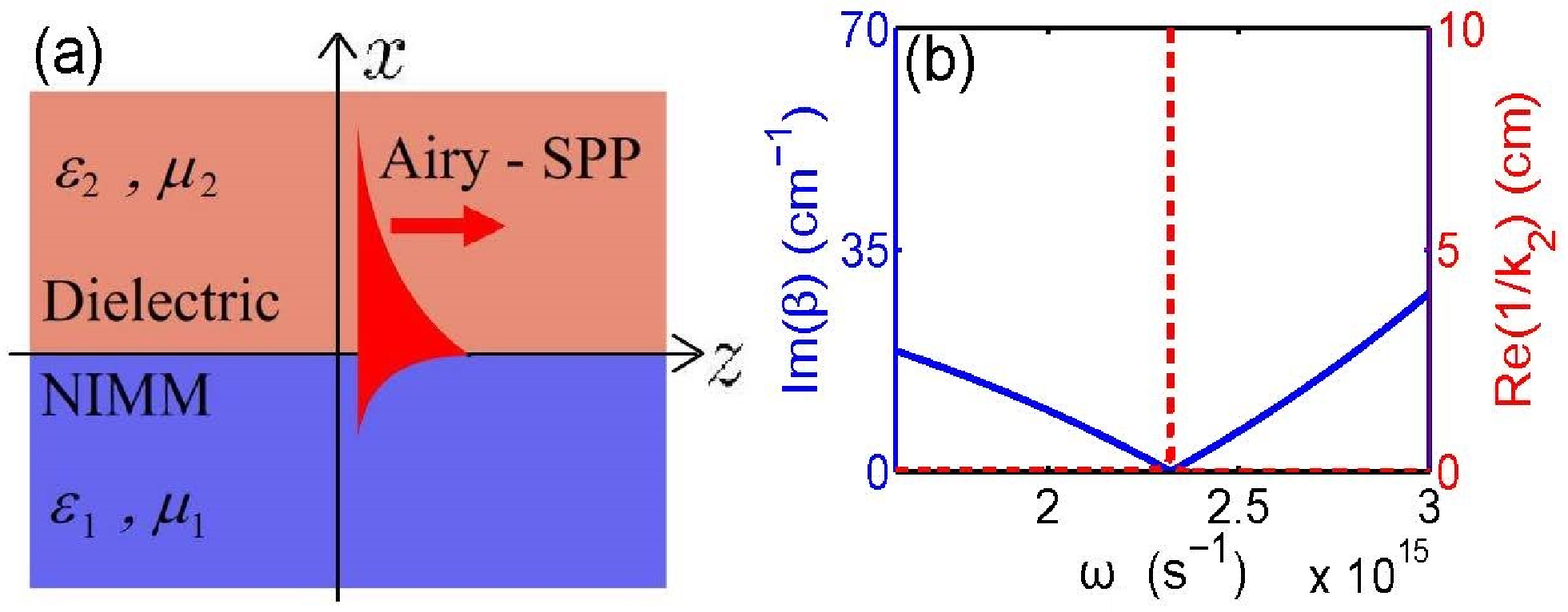}
\end{center}
\vskip 0.6cm \small \rm Fig.\hspace{0.1cm}1. (Color online) (a) The dielectric-NIMM interface. The NIMM (in the lower half-plane, $x<0$) has frequency-dependent permittivity
$\varepsilon_1$ and permeability $\mu_1$. The dielectric (in the upper half-plane, $x>0$) has frequency-independent permittivity $\varepsilon_2$ and permeability $\mu_2$.
SPP propagates along $z$-direction. (b) Im$(\beta)$ (blue solid line) and Re($1/k_2$) (red dashed line) as functions of oscillating frequency $\omega$.
\vskip 0.8cm \noindent \normalsize

\vskip 0.75cm

We choose Drude model to describe the electric permittivity and magnetic permeability
of the NIMM [19], i.e. $\varepsilon_{1}^{\rm NIMM}(\omega)=\varepsilon_{\infty}-\omega_e^2/(\omega^2+i\gamma_e\omega)$ and $\mu_{1}^{\rm NIMM}(\omega)=\mu_{\infty}-\omega_m^2/(\omega^2+i\gamma_m\omega)$,
where $\omega_e$ and $\omega_m$ are electric and magnetic plasmon frequencies, $\gamma_e$ and $\gamma_m$ are corresponding decay rates, and $\varepsilon_{\infty}$ and $\mu_{\infty}$ are background constants. For comparison, the parameters for conventional metals are given by $\varepsilon_{\infty}\approx1$ [i.e. $\varepsilon_{1}^{\rm metal}(\omega)=1-\omega_e^2/(\omega^2+i\gamma_e\omega)$] and $\mu_{1}^{\rm metal}(\omega)=1$.

\vskip 0.75cm

The dielectric-NIMM interface supports both transverse electric and transverse magnetic modes. Here we focus only on the transverse magnetic modes. We assume that the SPP propagates in the positive $z$ direction and the $x$ direction is perpendicular to the interface. Then the form of electromagnetic field is taken as ${\bf E}_j({\bf r},t)={\bf E}_j({\bf r})e^{\pm k_jx+i(\beta z-\omega t)}$ (``+'' corresponds to $j=1$ for the NIMM, and  ``$-$'' corresponds to $j=2$ for the dielectric). Here ${\bf E}_j({\bf r})$ ($j=1,2$) are slowly-varying envelope functions and $k_j^2=\beta^2-k_0^2n_j^2$ (with $\beta$ being the propagation constant, $k_0=\omega/c$, and $n_j^2=\varepsilon_j\mu_j$) satisfies the boundary condition ${k_1}/{\varepsilon_1}=-{k_2}/{\varepsilon_2}$. It is easy to obtain the linear dispersion relation of the system
\begin{equation}
\beta(\omega)=\frac{\omega}{c}\sqrt{\frac{\varepsilon_1\varepsilon_2(\varepsilon_2\mu_1-\varepsilon_1\mu_2)}{\varepsilon_2^2-\varepsilon_1^2}}.
\label{characteristic dispersion relation}
\end{equation}

\vskip 0.75cm

The above model can be realized by using a silver-based NIMM, with the structure of Ag-Mg$\rm F_2$-Ag sandwich. The parameters are chosen as [19]: $\gamma_{e}=2.73\times10^{13}\,{s^{-1}}$, $\omega_{e}=1.37\times10^{16}\,{s^{-1}}$, $\gamma_{m}=10^{11}\,{\rm s}^{-1}$, and $\omega_{m}=0.2\omega_{e}$. The background constants are fixed as $\varepsilon_{\infty}=\mu_{\infty}=1.3$. (We shall also use Ag to replace the NIMM for comparison). The parameters for the dielectric are chosen as $\varepsilon_2=1.5$ and $\mu_2=1$. The SPP can be generated via optical grating coupling or prism coupling technique along the surface of the NIMM under the phase-matching condition.

\vskip 0.75cm

In Fig.~1(b), we show the imaginary part of propagation constant, Im($\beta$), based on Eq.~(\ref{characteristic dispersion relation}), and the real part of transverse confinement, Re$(1/k_2)$. We see that the Ohmic loss of the NIMM can be completely suppressed at a particular frequency $\omega_0=2\pi\times3.8197\times10^{14}\,{s^{-1}}$ where Im$(\beta)=0$. The physical reason for such suppression is due to the destructive interference between the electric- and magnetic-field absorptions of the NIMM because $\mu_1$ and $\varepsilon_1$ are both negative.

\vskip 0.75cm

However, the complete suppression of the Ohmic loss is unfortunately accompanied by a deconfinement of the SPP in the dielectric, i.e., Re$(1/k_2)\rightarrow\infty$ when Im$(\beta)\rightarrow0$. In order to obtain a significant suppression of the Ohmic loss and a required SPP confinement simultaneously, one must choose the optical frequency to have a small deviation from the lossless point $\omega_0$. Here we take $\omega=\omega_p=2\pi\times3.8462\times10^{14}\,{s^{-1}}$. As a result, a small absorption exists for the confined SPP.

\begin{center}
\includegraphics[width=0.8\columnwidth]{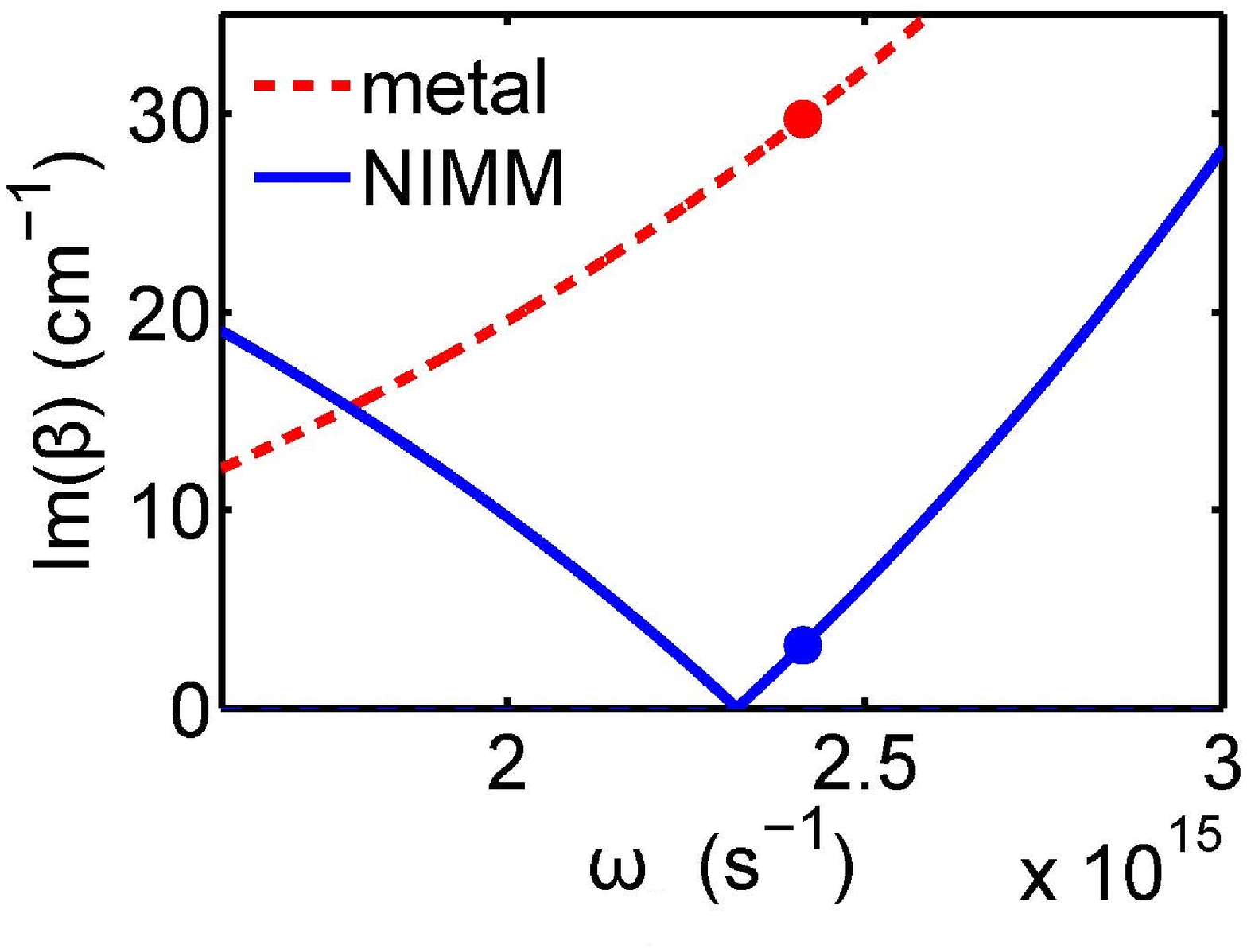}
\end{center}
\vskip 0.5cm \small \rm Fig.\hspace{0.1cm}2. (Color online) Im($\beta$) of the SPP as a function of $\omega$ at the dielectric-NIMM interface (blue solid line) and the dielectric-metal interface (red dashed line). Im$(\beta^{\rm NIMM})=3.16\,{\rm cm^{-1}}$ ({\color{red}blue dot}) and Im$(\beta^{\rm metal})=29.73\,{\rm cm^{-1}}$ ({\color{red}red dot}) at $\omega=\omega_p=2\pi\times3.8462\times10^{14}\,{s^{-1}}$.
\vskip 0.8cm \noindent \normalsize

\vskip 0.75cm

In Fig.~2, we show Im($\beta$) for a dielectric-NIMM interface and a dielectric-metal interface for comparison. We see that, different from the dielectric-NIMM interface, Im($\beta$) for the dielectric-metal interface has no zero point, i.e. there is no destructive interference. With the given parameters, we obtain that ${\rm Re}(1/k_{2}^{\rm NIMM})=1.90\times10^{-4}\,{\rm cm}$, $\beta^{\rm NIMM}=(9.86\times10^4+i3.16)\,{\rm cm^{-1}}$ ({\color{red} blue dot}), and $\beta^{\rm metal}=(0.10\times10^4+i29.73)\,{\rm cm^{-1}}$ ({\color{red} red dot}) at $\omega=\omega_p$. Consequently, the propagation loss of the dielectric-metal interface is much larger than that of the dielectric-NIMM interface.

\vskip 0.75cm

We now consider the propagation of an Airy SPP along the dielectric-NIMM interface.
We stress that full vector Maxwell's equations and related boundary conditions must be considered because the optical field is localized at the interface within a subwavelength
volume [20].  $E_{\alpha j}({\bf r})=(E_{xj}({\bf r}),E_{yj}({\bf r}),E_{zj}({\bf r})\,)$ ($j=1,2$)
satisfy the
Helmholtz equation $\nabla\times\left(\nabla\times{\bf E}_j\right)-n_j^2k_0^2{\bf E}_j=0$, i.e.
\begin{subequations}
\begin{eqnarray}
\frac{\partial^2E_{yj}}{\partial x\partial y}-\frac{\partial^2E_{xj}}{\partial y^2}-\frac{\partial^2E_{xj}}{\partial z^2}+\frac{\partial^2E_{zj}}{\partial z\partial x}=n_j^2k_0^2E_{xj},\label{equation for Ex}\\
\frac{\partial^2E_{zj}}{\partial y\partial z}-\frac{\partial^2E_{yj}}{\partial z^2}-\frac{\partial^2E_{yj}}{\partial x^2}+\frac{\partial^2E_{xj}}{\partial x\partial y}=n_j^2k_0^2E_{yj},\label{equation for Ey}\\
\frac{\partial^2E_{xj}}{\partial z\partial x}-\frac{\partial^2E_{zj}}{\partial x^2}-\frac{\partial^2E_{zj}}{\partial y^2}+\frac{\partial^2E_{yj}}{\partial y\partial z}=n_j^2k_0^2E_{zj}.\label{equation for Ez}
\end{eqnarray}
\end{subequations}

\vskip 0.75cm

The first step to get Airy SPP solutions is to derive the envelope equation for SPP propagation,
which can be obtained by employing a method of multiple scales [19]. To this end, we consider a group of multi-scale variables $x_0=x$, $y_1=\epsilon y$, and $z_{j}=\epsilon^{j}z\,(j=0,2)$, where $\epsilon$ is a dimensionless small parameter. In addition, we take the following asymptotic expansions $E_{\alpha j}({\bf r})=[\epsilon E^{(1)}_{\alpha j}+\epsilon^{3}E^{(3)}_{\alpha j}+\cdots]\exp(i\beta z_0),\,(\alpha=x,z)$, and
$ E_{yj}({\bf r})=[\epsilon^2 E^{(2)}_{yj}+\epsilon^{4}E^{(4)}_{yj}+\cdots]\exp(i\beta z_0)$,
with all quantities on the right hand side in the square bracket being considered as functions of multi-scale
variables $x_0$, $y_1$, and $z_2$.

\vskip 0.75cm

At the $\epsilon$-order, we obtain the equation
$\partial^2E^{(1)}_{zj}/\partial x_0^2-k_j^2E^{(1)}_{zj}=0$,
which admits the solution
$E^{(1)}_{zj}=A^{(1)}_je^{\pm k_{j}x_0}$, with the envelope $A^{(1)}_j=A^{(1)}_j(y_1,z_2)$ dependent on the slow variables due to the modulation by the dielectric. In addition, $E^{(1)}_{xj}=\mp (i\beta/k_j)A^{(1)}_je^{\pm k_{j}x}$. From the boundary condition $E^{(1)}_{z1}=E^{(1)}_{z2}$, one can get $A_1^{(1)}=A_2^{(1)}=A$. The solution of $E^{(2)}_{yj}$ can be obtained at the $\epsilon^2$-order, which is skipped here for saving space.

\vskip 0.75cm

At the $\epsilon^3$-order, we obtain the equation
$\partial^2E^{(3)}_{zj}/\partial x_0^2-k_j^2E^{(3)}_{zj}=L_{zj}e^{\pm k_jx}$,
with $L_{zj}=-2i\beta(\partial A_j^{(1)}/\partial z_2)-\partial^2 A_j^{(1)}/\partial y_1^2$. A solvability condition at this order requires that $L_{zj}=0$, which can be written into a dimensionless form, reading as
$2i(\partial u/\partial s)+\partial^2 u/\partial \eta^2=0$.
Here $s=z/(\beta l_0^2)$, $\eta=y\l_0$, and $u=A/U_0$, with
$l_0$ ($U_0$) the typical width (amplitude) of the light field. Note that in the above derivation
the imaginary part of $\beta$ is neglected because it is much smaller than the corresponding real part (the small imaginary part will be considered in the numerical simulation below). The equation for $u$ admits a finite-power Airy function solution [2]:
$u(s,\eta;a)={\rm Ai}(\eta-{s^2}/{4}+ias)\exp[i{s}
(\eta-{s^2}/{6})/{2}]\exp[a\eta-{as^2}/{2}+i{a^2s}/{2}]$.
Here $a$ is a small positive parameter introduced to avoid the divergence of the light power of the beam.
The initial condition is $u(s=0,\eta;a)={\rm Ai}(\eta)\exp(a\eta)$, which can be generated by using a
spatial light modulator.

\vskip 0.75cm

With the solution given above, when returning to the original variables
we obtain the Airy SPP of the system
\begin{eqnarray}
\nonumber {\bf E}({\bf r},t)&=&U_0\left({\bf e}_x+{\bf e}_y\frac{i}{\beta}\frac{\partial}{\partial y}\mp{\bf e}_z\frac{i\beta}{k_j}\right)e^{i(\beta z-\omega t)}\notag\\
&&\times{\rm Ai}\left(\frac{y}{l_0}-\frac{z^2}{4\beta^2 l_0^4}+ia\frac{z}{\beta l_0^2}\right)e^{i\frac{z}{2\beta l_0^2}\left(\frac{y}{l_0}-\frac{z^2}{6\beta^2 l_0^4}\right)}\notag\\
&&\times e^{a\frac{y}{l_0}-\frac{az^2}{2\beta^2l_0^4}+i\frac{a^2z}{2\beta l_0^2}}.
\label{explicit linear solution}
\end{eqnarray}
We see that the Airy SPP generated in the dielectric-NIMM interface propagates along a parabolic
trajectory $y=z^2/(4\beta^2l_0^3)$ and preserves their wave shape during propagation.
\begin{center}
\includegraphics[width=\columnwidth]{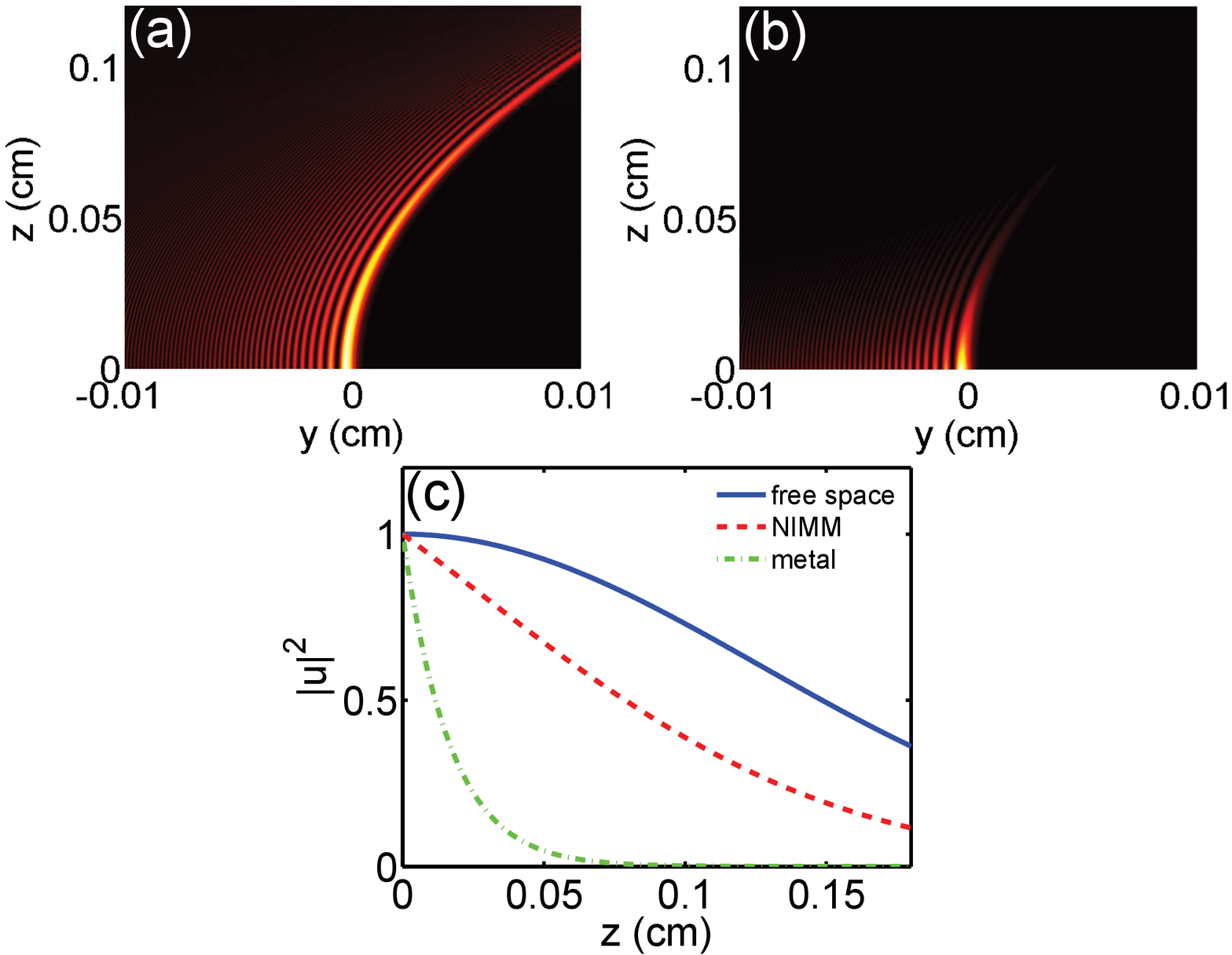}
\end{center}
\vskip 0.5cm \small \rm Fig.\hspace{0.1cm}3. (Color online) Intensity profiles of the Airy-type SPPs at dielectric-NIMM interface (a) and dielectric-metal interface (b), respectively. (c) Maximum intensity max$(|u|^2)$ as a function of $z$ in free space (blue solid line), dielectric-NIMM interface (red dashed line), and dielectric-metal interface (green dash-dotted line). The parameters are taken as $l_0=3\,{\rm \mu m}$ and $a=0.005$. Other parameters are the same with those used in Fig.~2.
\vskip 0.8cm \noindent \normalsize

\vskip 0.75cm

Shown in Fig.~3(a) and Fig.~3(b) are intensity profiles of the Airy SPPs at dielectric-NIMM interface and dielectric-metal interface, respectively. The parameters are taken as $l_0=3\,{\rm \mu m}$ and $a=0.005$ with other parameters being the same with those used in Fig.~2. We see that at the dielectric-NIMM interface the Airy SPP can propagate to a long distance without obvious attenuation. This is because the absorption of the Airy SPP due to the Ohmic loss of the NIMM is largely suppressed. However, at the dielectric-metal interface the Airy SPP propagates only for a short distance and attenuates due to the significant Ohmic loss. To see this more clearly, we define the absorption length of the Airy SPP as $L_{\rm abs}\equiv1/[2{\rm Im}(\beta)]$. From the above results we obtain
\begin{equation}
L_{\rm abs}^{\rm NIMM}=0.1043\, {\rm cm}, \qquad L_{\rm abs}^{\rm metal}=0.0167\, {\rm cm}.
\end{equation}
Thus, the absorption length of the dielectric-NIMM interface is more than 6-time large than that of the dielectric-metal interface, which means that the Airy SPP obtained in the dielectric-NIMM interface may propagate more than 6-time long distance than that in the dielectric-metal interface.

\vskip 0.75cm

Fig.~3(c) shows the maximum intensity, max$(|u|^2)$, as a function of the propagating distance $z$ in free space, dielectric-NIMM interface, and dielectric-metal interface. We see that the curves of the free space and the dielectric-NIMM interface are very close, indicating that there is only slight absorption in the dielectric-NIMM interface. However, the dielectric-metal interface exhibits a significant absorption.

\vskip 0.75cm

In conclusion, we have proposed a scheme to obtain a low-loss propagation of Airy SPPs along the dielectric-NIMM interface. By using the transverse-magnetic mode and the related destructive interference effect between electric and magnetic absorption responses, we have shown that the propagation loss of the Airy SPPs can be largely suppressed when the optical frequency is close to the lossless point of the NIMM. As a result, the Airy SPPs obtained in our scheme can propagate more than 6-time long distance than that in conventional dielectric-metal interfaces.
The low-loss and non-diffraction characters of the Airy SPPs at the dielectric-NIMM interface may have promising applications in light and quantum information processing at nanoscale.

\vskip 0.95cm

This work was supported by the NSF-China under Grants No. 11174080, 11474099, and 11475063.

%%%%%%%%%%%%%%%%%%%%%%%%%%%%%%%%%%%%%%%%%%%%%%%5

\end{document}